# High field charge order across the phase diagram of YBa₂Cu₃Oᵧ

F. Laliberté[1], M. Frachet[1], S. Benhabib[1], B. Borgnic[1], T. Loew[2], J. Porras[2], M. Le Tacon[2,3], B. Keimer[2], S. Wiedmann[4], Cyril Proust[1] and D LeBoeuf[1]

[1] *Laboratoire National des Champs Magnétiques Intenses (LNCMI-EMFL), (CNRS-INSA-UGA-UPS), Toulouse / Grenoble, 31400/38042, France.*

[2] *Max-Planck-Institut für Festkörperforschung, Heisenbergstrasse 1, D-70569 Stuttgart, Germany.*

[3] *Karlsruher Institut für Technologie, Institut für Festkörperphysik, D-76344 Eggenstein-Leopoldshafen, Germany.*

[4] *High Field Magnet Laboratory (HFML-EMFL) and Institute for Molecules and Materials, Radboud University, Toernooiveld 7, 6525 ED Nijmegen, The Netherlands*

**In hole-doped cuprates there is now compelling evidence that inside the pseudogap phase, charge order breaks translational symmetry leading to a reconstruction of the Fermi surface. In YBa₂Cu₃Oᵧ charge order emerges in two steps: a 2D order found at zero field and at high temperature inside the pseudogap phase, and a 3D order that is superimposed below the superconducting transition $T_c$ when superconductivity is weakened by a magnetic field. Several issues still need to be addressed such as the effect of disorder, the relationship between those charge orders and their respective impact on the Fermi surface. Here, we report high magnetic field sound velocity measurements of the 3D charge order in underdoped YBa₂Cu₃Oᵧ in a large doping range. We found that the 3D charge order exists over the same doping range as its 2D counterpart, indicating an intimate connection between the two distinct orders. Moreover, we argue that the Fermi surface is reconstructed above the onset temperature of 3D charge order.**

The discovery of charge order in the pseudogap phase of cuprates has triggered a lot of experimental[1-5] and theoretical[6-11] works. The relation between charge order and pseudogap is not yet settled but some models predict that they are intertwined: in the spin fermions model[8-10] the pseudogap can for instance be understood as a superposition of $d$-wave superconductivity and a quadrupole-density wave. In other scenarios the pseudogap phase corresponds to a nematic phase precursor to charge ordering at lower temperature[6,7] or to a fluctuating charge order[12,13]. Two distinct charge orders have been detected in YBa₂Cu₃Oᵧ (YBCO). First a two-dimensional (2D) short range (but static) bidirectional charge density wave (CDW) appears at high temperature between $T_c$ and the pseudogap temperature $T^*$. Comprehensive X-ray measurements[14,15] in YBCO have shown that the 2D CDW is incommensurate and occurs in the doping range $p \approx 0.08$ to $p \approx 0.16$. The propagation vectors along $a$ and $b$ directions are $Q_a = (\delta_a, 0, 0.5)$ and $Q_b = (0, \delta_b, 0.5)$ where $\delta_{a,b} \approx 0.3 - 0.34$ with in-plane correlation lengths that are at most 20 unit cells and weak anti-phase correlation between neighboring bilayers. It has been detected later by NMR measurements[16] as a broadening of the quadrupolar NMR lines, which has been interpreted as a signature of an inhomogeneity of the charge distribution. The second CDW is a three-dimensional (3D) ordered state with in-plane CDW modulations along the $b$ directions only –though at the same value than the 2D CDW–, that appears below $T_c$ and above a threshold field[2,17]. Moreover, the periodicity along $c$-axis is close to unity, meaning that the 3D CDW modulation is in-phase in neighbouring bilayers. Compared to the 2D short range CDW, the in-plane and $c$-axis correlation lengths are greatly enhanced. The 3D CDW has been first observed in YBCO using high field NMR measurements. It has been characterized by recent X-ray measurements[18-20] in high fields at two doping level, $p \approx 0.11$ and $p \approx 0.12$ corresponding to oxygen order ortho II (O-II) and ortho VIII (O-VIII), respectively. It gives rise to anomaly in the field dependence of the elastic constants via sound velocity measurements[21] and of the magnetization[22], the only thermodynamic probes that so far have detected a signature of a phase transition to the CDW state. It has been established by NMR[16] and X-ray[18-20] measurements that these charge orders coexist at low temperatures but are related since they share the same periodicity along $b$. The co-existence of two distinct charge orders, with different boundaries in the temperature-doping phase diagram raises a certain number of interesting questions: Do both charge orders share the same critical doping? What is their respective impact on the Fermi surface? What is the role of disorder in their occurrence?

Here we report a full doping dependence of high field sound velocity measurements, which probe the 3D charge order at low temperatures. Anomalies are now seen in the temperature dependence of sound velocity measured at high field showing a clear negative jump as expected for a $2^{nd}$ order phase transition. From the doping dependence of the threshold field $H_{CO}$ and of the onset temperature $T_{CO}$ of the 3D charge order, we find that the two charge orders occur in the same doping range. Moreover, the significant difference observed between the temperature of the onset of 3D charge order and of the sign change in the Hall coefficient shows that Fermi surface reconstruction occurs before the 3D charge order sets in. This is corroborated by comparing the de Broglie wavelength and the correlation length of the 2D charge order.

## Results

The sound velocity is defined as $v_s = \sqrt{\dfrac{c_{ij}}{\rho}}$ for propagation directions along high symmetry axis, where $\rho$ is the density of the material, $c_{ij} = \dfrac{1}{V}\dfrac{\partial^2 F}{\partial \varepsilon_i \partial \varepsilon_j}$, $F$ the free energy and $\varepsilon_i$ the strain . Changes in the elastic constants $c_{ij}$ are expected whenever a strain dependent phase transition occurs[23]. This is the case at the superconducting transition where a negative jump of the elastic constant is seen at $T_c$ (see Supplementary Fig. S1). The amplitude of the jump of the elastic constants and of the heat capacity $\Delta C_p(T_c)$ are proportional through the Ehrenfest relation: $\Delta c_{ii}(T_c) = -\dfrac{\Delta C_p(T_c)}{T_c}\left(\dfrac{dT_c}{d\varepsilon_i}\right)^2$, where the minus sign explains the downward jump in the elastic constant (see discussion in the Supplementary). This feature is also seen at the onset temperature $T_{CO}$ of the 3D charge order. Indeed, in Fig. 1 we show the in-field temperature dependence of the sound velocity (background subtracted, see Fig. S2 for raw data and fit) in YBCO at three different doping levels, for which $H_{c2} < 30$ T (ref. [24]) such that the normal state is reached down to low temperatures. This feature is the first evidence of a $2^{nd}$ order phase transition-like anomaly in the temperature dependence of a thermodynamic probe at $T_{CO}$. The size of the anomaly is similar for $p = 0.106$ and $p = 0.110$ but it gets smaller for $p = 0.122$.

Fig. 2a and Fig. 2b show the field dependence up to 35 T of the sound velocity at different temperatures in YBCO at $p = 0.122$. For magnetic field below the irreversibility field, there is a strong contribution from the vortex lattice (see discussion in the Supplementary and Fig. S3). At low fields in the pinned solid vortex phase, the vortex lattice contribution leads to an increase of the sound velocity. This vortex lattice contribution is lost when vortices are depinned leading to a large drop in the sound velocity. The midpoint of this step-like transition associated with vortex depinning is labeled $H_v$. At field above $H_v$, the weakening of superconductivity with increasing magnetic field induces a progressive decrease of the sound velocity, until a pronounced hardening of the elastic constant that we attribute to the 3D charge order[21]. It leads to a minimum at a field that we define as the threshold charge order field, $H_{CO}$(T). By performing field sweeps at different temperatures, we can track $H_{CO}$(T) versus T and draw up the phase diagram shown in Fig. 2c. Black squares and red circles correspond to $H_v(T)$ and $H_{CO}(T)$, respectively. The phase diagram can be interpreted as follow: at high fields, in the normal state, the charge order transition is field independent. Due to competition effect when lowering the field close to $H_{c2}$, superconductivity impedes charge order to appear, pushing the vertical phase boundary to go almost horizontal such that 3D charge order only appears at finite magnetic field at low $T$. Similar phase diagram has been obtained by high field X-ray[19] and Seebeck coefficient[25] measurements. Based on this phase diagram, we define two quantities: $H_{CO} = H_{CO}(T\rightarrow0)$ and $T_{CO}$ which corresponds to the vertical line in Fig 2c where 3D charge order sets in for $H > H_{c2}$. Several measurements have been performed in the doping range between $p = 0.071$ and $p = 0.154$ and Fig. 3 shows the field dependence of the sound velocity at $T = 20$ K at all doping. We use the minimum in $\Delta v_s/v_s(H)$ above $H_v$ as a criterion to pinpoint the 3D charge order $H_{CO}$ and extract its doping dependence in the doping range $p = 0.095$ to $p = 0.139$ (see Fig. 4). It is worth noting that the field dependence of the sound velocity above $H_v$ at $p = 0.071$ (see curves at different temperatures in Fig. S3b) and $p = 0.154$ is featureless, with almost no field dependence. In fig. S5c, we show the field dependence for $p = 0.154$ of the $c_{66}$ mode at $T = 4.2$ K, for which there is no sign of transition up to 66 T. We conclude that there is no 3D CDW at these doping levels. The full set of data at all doping along with the temperature – magnetic field phase diagrams deduced from the measurements are shown in the Supplementary Fig. S4 and Fig. S5. The overall shape of these phase diagrams are in agreement with models based on competition between CDW and superconductivity, as discussed below.

## Discussion

Indeed, a theory based on a phenomenological nonlinear sigma model which formulates the competition between CDW and fluctuating superconductivity[26] gives qualitative agreement with the experiment if the effect of disorder is taken into account[27,28]. In the model of ref. [28] where the effect of magnetic field is incorporated, the red symbols in Fig. 2c represent a crossover between a short-range 2D CDW order to a long-range 3D CDW at high fields. Even if theory precludes long range incommensurate CDW order in disordered systems, the notion of 'failed thermodynamic transition' has been used to describe the anomaly observed at the temperature / magnetic field where the mean field theory predicts thermodynamic transition. As the disorder increases, one expects the transition to become more gradual and it leads to crossover. In YBCO, the main source of disorder come from the chains that has an impact in the $CuO_2$ planes. Indeed, oxygen disorder in the chain layer creates point-like defects (the ends of finite length chainlets) as well as domain walls caused by phase slips in the chain ordering pattern[29]. Since the correlation length of the chain order is higher at O-II doping compare to O-VIII and O-III doping[30], it can explain the pronounced anomaly in the temperature dependence of the sound velocity at $T_{CO}$ around O-II doping (see Fig. 1). Another effect that needs certainly to be taken into account is the change in the $c$-axis correlation length of the CuO chain superstructure order which is finite at O-II doping while the other CuO chain superstructures are mainly 2D. It is conceivable that disorder in the chain structure along the $c$-axis for oxygen content away from O-II could smear out the transition towards the 3D ordered CDW. While the non-linear sigma model succeeds to explain most of the salient feature related to the 2D / 3D CDW, one must admit that the situation is more complex in YBCO since it has been found that the 3D ordered CDW at high fields develops on top of the 2D CDW and it is not simply a crossover, that is to say they coexist at low temperature in presence of a magnetic field[31]. In addition, the two charge orders differ in the sense that the 3D CDW is 3D ordered but uniaxial along $b$ in the plane (although with the same wavevector than the 2D CDW).

From the temperature sweeps shown in Fig. 1 and the phase diagrams at different doping levels, we were able to compile the doping dependence of the onset field $H_{CO}$ and the onset temperature $T_{CO}$ of the 3D ordered charge order shown in Fig. 4 and Fig. 5, respectively. In Fig. 4a, the doping dependence of $H_{CO}$ shows that it is systematically lower than $H_{c2}$. Fig. 4b shows that in the limit $T \rightarrow 0$, 3D charge order always appears around $H_{CO} \approx 0.8\,H_{c2}$, despite the pronounced doping dependence of the ratio $T_{CO}$ / $T_c$. Fig. 4a reveals also that NMR gives lower $H_{CO}$ than sound velocity measurements. One way to understand this effect is to assume that 3D charge order appears first confined inside vortex cores and observed by NMR due to its sensitivity to normal quasiparticles within the vortex cores[17]. The observation of in-plane precursor correlations of the 3D charge order[19] can also explain this difference. NMR would be sensitive to this precursor phase but not sound velocity since it is a thermodynamic probe and hence is mostly sensitive to the onset of a 3D long range, static order parameter. $H_{CO}$ measured by sound velocity is indeed in better agreement with thermal Hall effect[32], X-ray[18-20] and recent Seebeck coefficient[25] measurements in finite magnetic field. In Fig. 5 we plot the onset temperatures $T_{CO}$ for the 3D ordered CDW. The comparison of $T_{CO}$ with the 2D CDW onset temperatures deduced from X-ray measurements[14,15] leads to our first main finding: 2D and 3D CDW occur in the same doping range, within experimental accuracy. In other words, the charge orders are intimately linked even though the situation does not correspond to a crossover between the two. Note that the critical point of pseudogap ($p^* \sim 0.19$ from ref. [33]) and of CDW are distinct and well separated. But it raises the question: why both charge orders occur only in such limited doping range? In YBCO spin density wave (SDW) occurs for $p < 0.08$ and the competition between SDW and CDW (ref. 34) can explain the location of the critical point at $p \approx 0.08$. On the right side of the CDW dome ($p \approx 0.16$), the proximity of the critical point of the pseudogap ($p = 0.19$) leads to a change in the carrier density[35], thus in the Fermi surface topology that can be detrimental for charge order.

Finally, we discuss the implication of our finding for the Fermi surface reconstruction process induced by charge order. Since the correlation lengths are strongly enhanced at low temperature and under magnetic field[18-20], it would be natural to assume that the 3D ordered charge order is responsible for the Fermi surface reconstruction and that the 2D CDW is a consequence of the presence of disorder[36,37]. Fermi surface reconstruction by an uniaxial charge order cannot produce an electron pocket except if a nematic phase exists at high temperature that distorts the original Fermi surface[38]. In the tetragonal system like $HgBa_2CuO_{4+\delta}$ (Hg1201), a negative Hall effect[39] and quantum oscillations[40,41] with similar frequency as in YBCO have been observed suggesting a similar phenomenology as in YBCO. While a nematic phase might be hidden in Hg1201 due to its tetragonal crystal symmetry giving rise to equally populated domains along both in-plane directions, only short-range charge ordering has been reported so far for this compound[5].

In Fig. 5 we compare the onset temperature $T_{CO}$ of the 3D CDW deduced from ultrasound and NMR measurements[17] with $T_0$, the temperature where the Hall effect changes sign[42]. The latter is a signature of the presence of an electron pocket in the reconstructed Fermi surface at low temperature. For $p > 0.11$, the onset of 3D CDW occurs at a temperature where the Hall effect is already negative. In Fig. 6a we compare the in-field temperature dependence of the sound velocity and of the Hall effect for $p = 0.12$, which clearly indicates that Fermi surface reconstruction takes place at higher temperature than the onset of 3D charge order. It opens another question: What is the shortest correlation length that can induce a Fermi surface reconstruction? At high temperatures, where the mean-free-path is short, quasiparticles can be sensitive even to short-range charge order. The thermal de Broglie wavelength $\xi_{th} = \hbar v_F / k_B T$ should be compared with the correlation length of the 2D charge order $\xi_{CO}$. If $\xi_{CO} \geq \xi_{th}$ quasiparticles become sensitive to the charge order correlations, which might lead to a Fermi surface reconstruction. The average Fermi velocity $v_F$ can be deduced from quantum oscillations[43], $F = 540$ T and $m^* = 1.8 \, m_e$ leads to $v_F = 8.2 \, 10^4$ m s$^{-1}$. In Fig. 6b, we show the temperature dependence of $\xi_{th}$ and of the 2D charge order correlation length[14] measured in O-II YBCO in zero field. The de Broglie wavelength follows closely the correlation length of the charge order above $T_c$ and up to the temperature where the 2D charge order appears. This strongly suggests that 2D short range charge order can reconstruct the Fermi surface at high temperatures.

At low temperatures, however, the situation is less clear. In Fig. S6 we compare the quasiparticle mean free path deduced from quantum oscillations experiments with $\xi_{CO}(T = T_c)$, at different doping in YBCO. We see that, except for well-ordered YBCO O-II ($p \sim 0.11$), the two length scales are comparable, suggesting that quasiparticle mean free path is bound by the 2D charge order correlation length. Note that the two length scales might be limited due to a significantly enhanced quasiparticle scattering originating in the shorter correlation lengths of their CuO-chain ordering[30,44]. However, for YBCO O-II the mean free path is much longer than the 2D charge order correlation length and is comparable to the 3D charge order correlation length. Given that the wavevector of the modulation is close to a commensurate number ($\delta_b \approx 0.33$ from ref [20]), it is tempting to invoke a 'lock-in' transition at high fields at this particular oxygen ordering, leading to a strong enhancement of the correlation length of both charge order. Note that if the Fermi surface is already reconstructed by the 2D charge order at high temperature, the occurrence of the 3D ordered charge order with the same vector along the $b$ axis would not strongly affect the reconstructed Fermi surface.

This overall scenario would explain the negative Hall and Seebeck coefficients in Hg1201(ref. 39) and $La_{2-x}Sr_xCuO_4$ (ref. 45) where only short range charge orders have been observed[5,46] and long range charge order has been missing so far.

**Acknowledgments** This work was performed at the HMFL and the LNCMI, members of the European Magnetic Field Laboratory (EMFL). C.P. acknowledges funding from the French ANR SUPERFIELD, the Laboratoire d'Excellence NEXT. D.L. acknowledges funding from the French ANR UNESCOS, the Laboratoire d'Excellence LANEF and the Université Grenoble-Alpes (SMIng-AGIR).

**Competing Interests** The authors declare that they have no competing financial interests.

**Author Contributions** J. P., T. L., M. L. T. and B. K. grew, prepared (annealing, de-twinning) and characterized the samples. F. L., S. B., M. F., B. B., C. P., D. L. performed the high magnetic field measurements. S. W. provided instrumental support for the d.c. field experiment at HMFL. F. L., S. B., M. F. and D. L. carried out the data analysis. C. P. and D. L. wrote the manuscript with inputs from all co-authors.

Correspondence and requests for materials should be addressed to C. P. (cyril.proust@lncmi.cnrs.fr) and D. L. (david.leboeuf@lncmi.cnrs.fr).

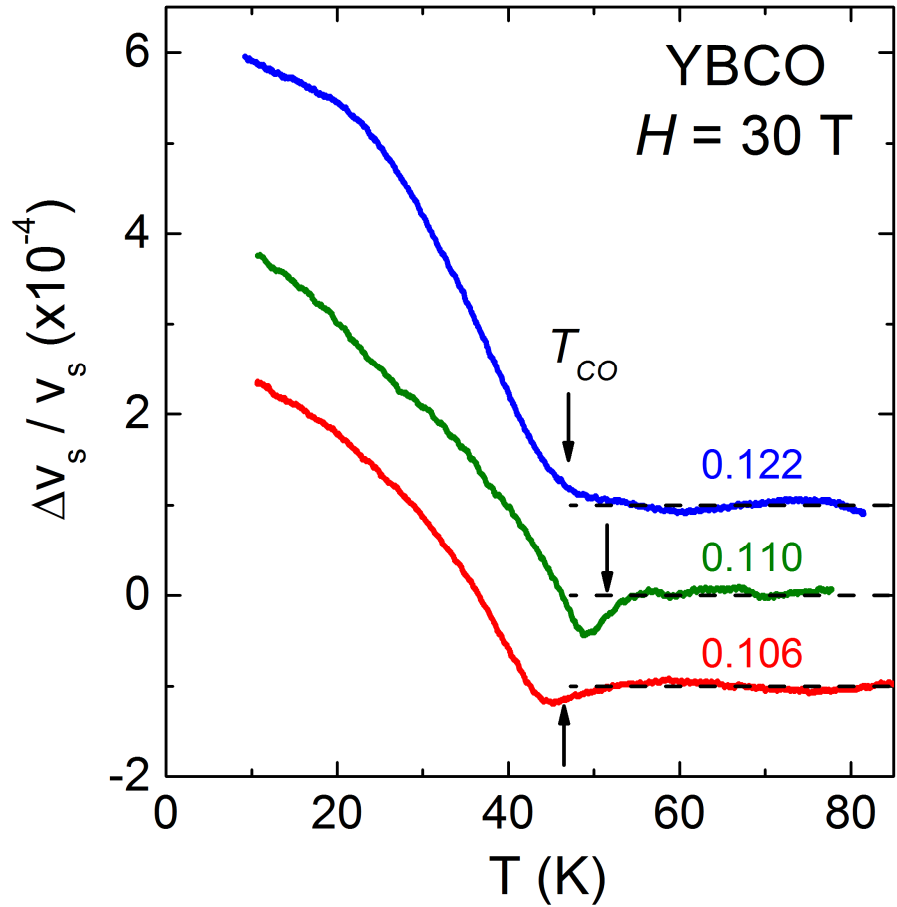

**Figure 1 | Charge order anomaly seen in the temperature dependence of the sound velocity in YBCO**

Sound velocity variation of the $c_{22}$ mode as a function of temperature measured in YBCO at $p = 0.106$ (red), $p = 0.110$ (green) and $p = 0.122$ (blue) at $H = 30$ T. A lattice background contribution has been subtracted as discussed in the Supplementary. The arrows indicate the charge order transition temperature $T_{CO}$. Curves are shifted vertically for clarity.

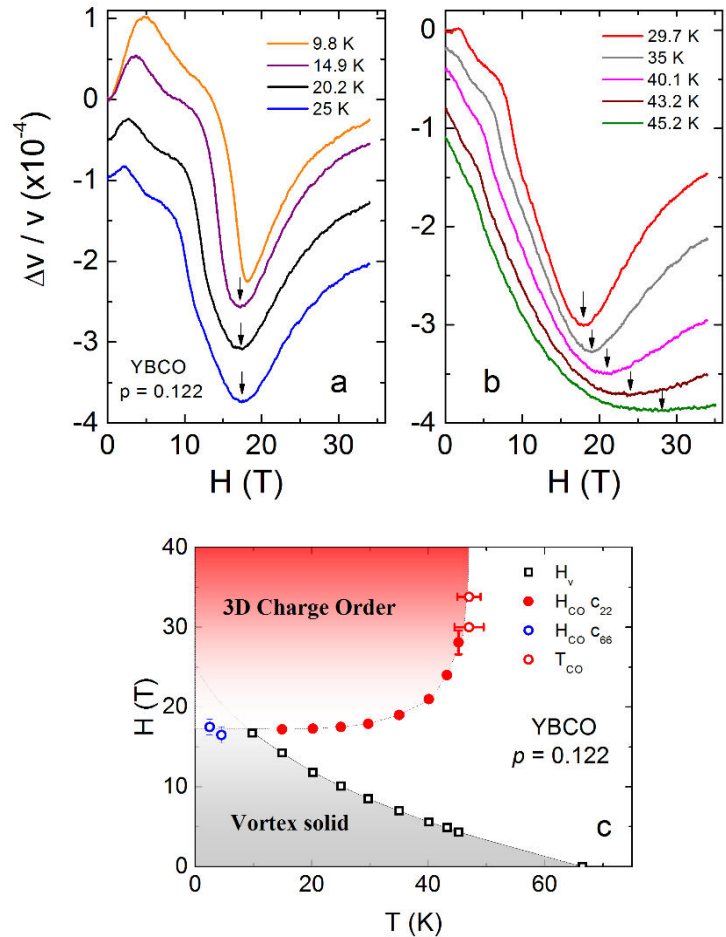

**Figure 2 | Temperature – magnetic field phase diagram of the charge order in YBCO at p = 0.12**

(a) and (b) Field dependence of the sound velocity of the $c_{22}$ mode in YBCO ($p$ = 0.122) measured at temperatures ranging from 9.8 K to 45.2 K. The arrows indicate $H_{CO}$, the charge order transition field. Curves are shifted vertically for clarity. (c): Temperature – magnetic field phase diagram of YBCO ($p$ =0.122) deduced from sound velocity measurements. $H_{CO}$ (full red circles) and $H_v$ (open black squares) are obtained from measurements shown in the upper panels. $T_{CO}$ (open red circles) is deduced from the temperature dependence of the sound velocity shown in Fig.1 and Fig. S2. Blue circles correspond to $H_{CO}$ measured at low temperatures using the field dependence of the transverse mode $c_{66}$ (see Fig. S5).

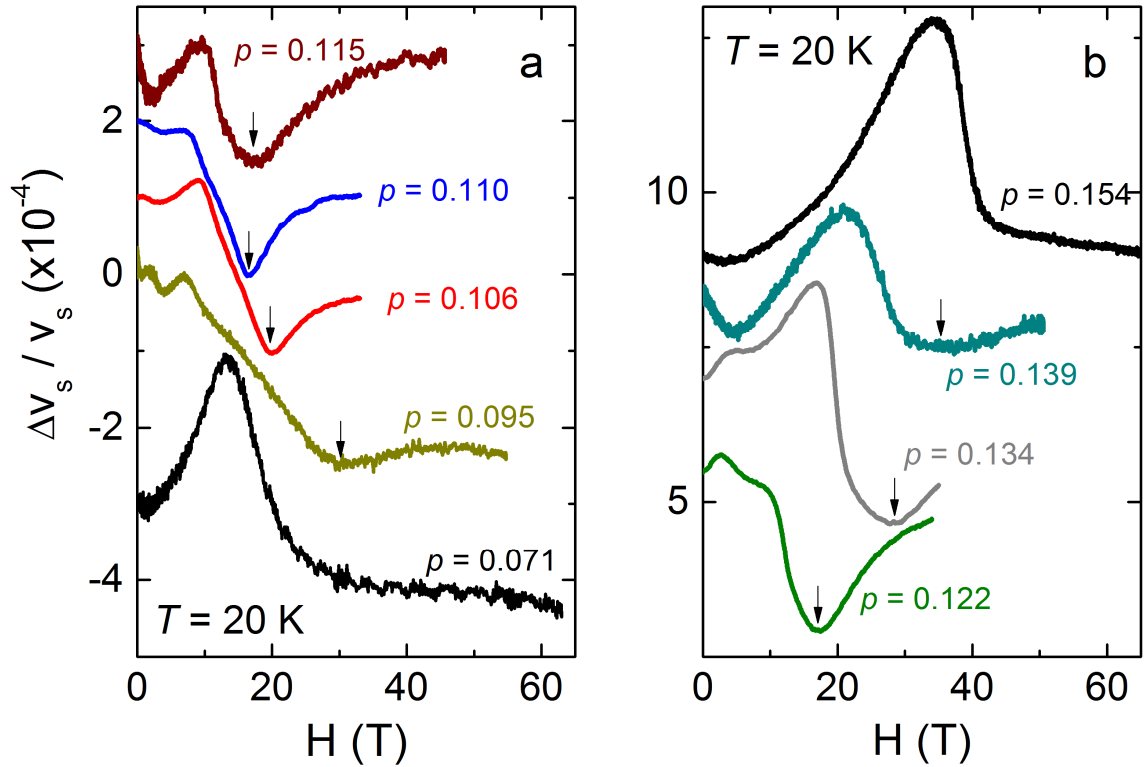

**Figure 3 | Charge order anomaly seen in the magnetic field dependence of the sound velocity in YBCO**

Field dependence of the sound velocity variation of the $c_{22}$ mode in YBCO for doping ranging from $p = 0.071$ to $p = 0.154$ measured at $T = 20$ K. Data for $p = 0.154$ are divided by a factor 5. Measurements were done either in DC fields up to 37.5 T or in pulsed fields. At $p = 0.071$ and $p = 0.154$, where no charge order transition is observed, the sound velocity first increases with magnetic field due to the vortex contribution. The large drop signals the loss of this vortex contribution. At still higher fields the sound velocity decreases further as superconductivity is being suppressed by the magnetic field. Those features are also seen at other doping, but a change of slope is observed due to the occurrence of charge order (arrows). Curves are shifted vertically for clarity.

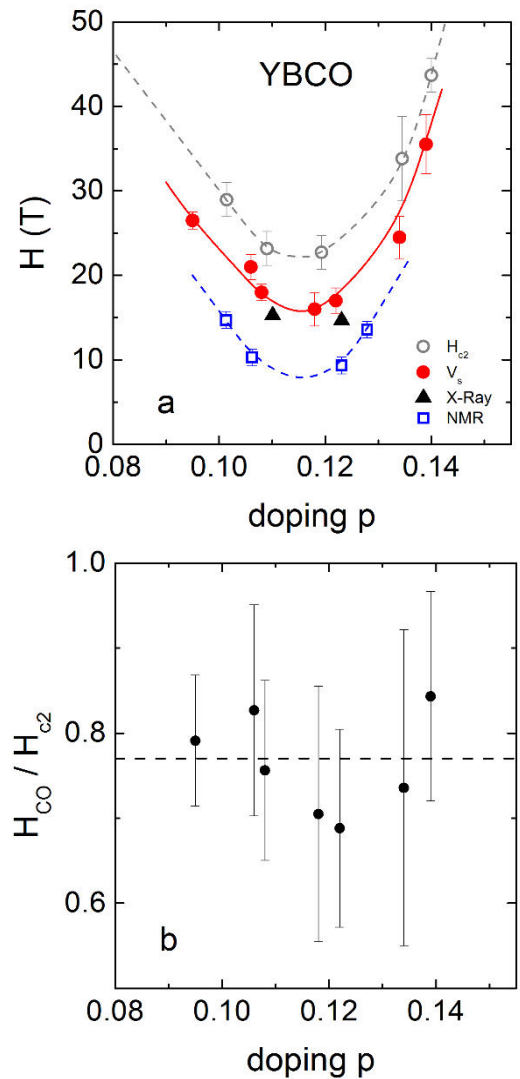

**Figure 4 |. Comparison of the threshold field $H_{CO}$ and $H_{c2}$ in YBCO**

(a) Magnetic field – doping phase diagram of YBCO comparing $H_{CO}$ and the upper critical field $H_{c2}$ at $T \rightarrow 0$ determined from thermal conductivity and magneto-resistance measurements[24]. $H_{CO}$ is determined by NMR[2,17] (blue squares), X-ray[19] (black triangles) and sound velocity (red full circles). Note the good agreement between X-ray and sound velocity data. Sound velocity data at $p = 0.108$ was obtained in ref. 21. (b) Doping dependence of the ratio $H_{CO}$ / $H_{c2}$, which appears to be almost constant over the entire doping range covered in this study.

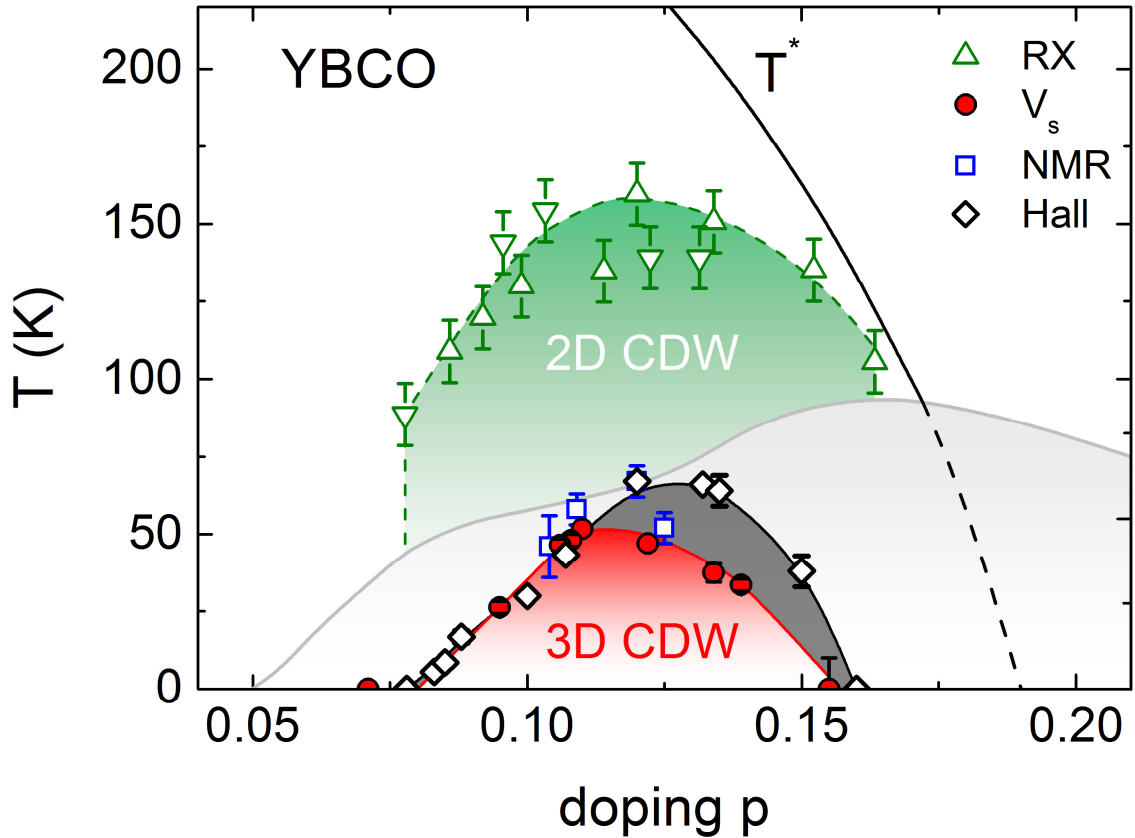

**Figure 5 |. Phase diagram of charge order in YBCO**

Temperature – doping phase diagram of charge orders in YBCO. X-Ray diffraction[15] (down green triangles) and Resonant X-Ray scattering[14] (up green triangles) give the onset temperature of 2D charge order in zero field. The onset temperature of 3D charge order in high fields is given by NMR[2,17] (blue squares) and sound velocity (red circles). Sound velocity data at $p$ = 0.108 was obtained in ref. 21. A comparison is made with $T_0$ (black diamonds) the temperature of the sign change of the Hall effect, a signature of an electron pocket in the reconstructed Fermi surface[35,42]. For $p$ > 0.11, the Fermi surface reconstruction takes place at higher temperature than the onset of 3D CDW. Solid black line is the pseudogap temperature $T^*$ in YBCO determined from the resistivity curvature map from Ref. 47. Dashed black line is a guide to the eye that extrapolates to $p^*$ = 0.19, the critical point of the pseudogap[33], that is distinct from the critical point of the CDW.

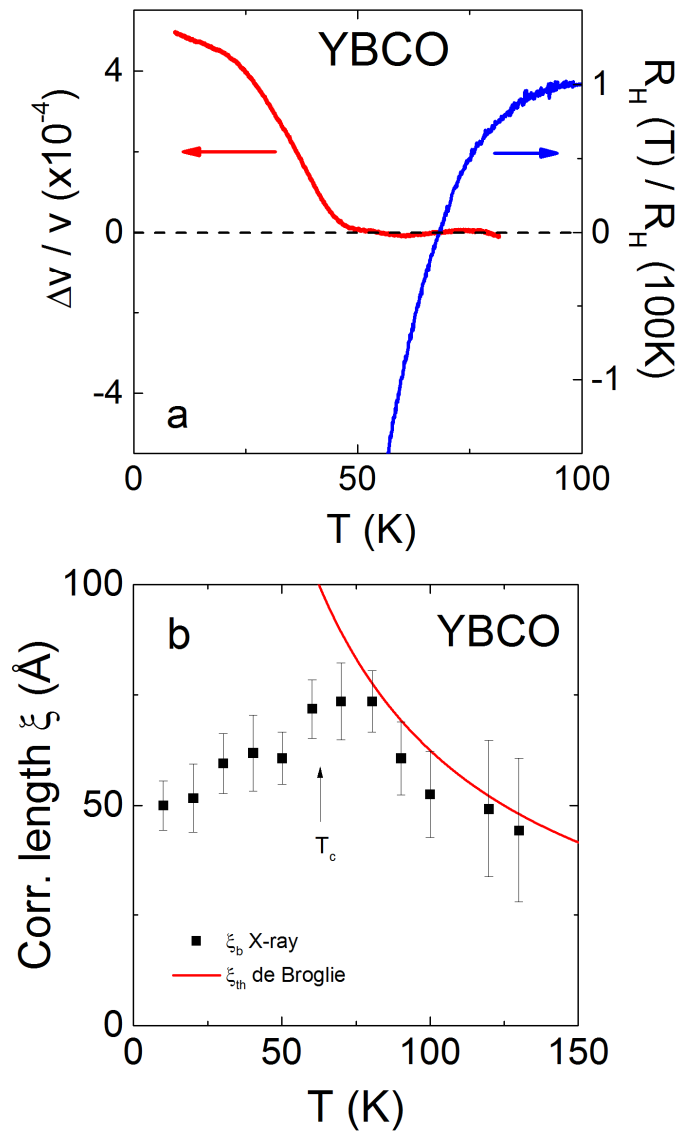

**Figure 6 |. Fermi surface reconstruction by 2D charge order**

(a) Comparison of the temperature dependence of the sound velocity induced by 3D charge order in YBCO at $p$ = 0.122 sample (red curve) and of the Hall coefficient in a YBCO sample at similar doping (blue curve). Note that the Hall coefficient changes sign 20 K above the onset of 3D charge order. (b) Temperature dependence of the correlation length of the 2D charge order[14] in YBCO O-II (black squares) and of the de Broglie wavelength (red line, see text).

# SUPPLEMENTARY INFORMATION

## 1. Methods

**SAMPLES**

The samples are detwinned single crystals of $YBa_2Cu_3O_y$ grown by the self-flux method; details on their preparation and characterization are given in Ref. [1-3].

The oxygen atoms in the CuO chains were made to order into the stable superstructure specific to the given oxygen concentration y. Samples with oxygen content $y=6.55$ and $y=6.51$ with ortho II order showed nice quantum oscillations in the sound velocity and attenuation (not shown), indicating high sample quality. The superconducting critical temperature $T_c$ were determined using the measurement of the elastic constant $c_{22}$ as shown in Fig. S1. The hole carrier concentration is deduced from the measurement of $T_c$ and the $c$-axis lattice parameter, using a relationship between $T_c$, $c$-axis lattice parameter and doping in the $CuO_2$ planes[4]. $T_c$, doping levels, and the structure of the CuO chains[5] for the different samples are listed in Table S1.

| y | $T_c$ (K) | Doping, $p$ (holes/Cu) | Structure |
|---|---|---|---|
| 6.45 | 34.0 | 0.071 | O-II |
| 6.48 | 55.7 | 0.095 | O-II |
| 6.51 | 60.0 | 0.106 | O-II |
| 6.55 | 62.0 | 0.110 | O-II |
| 6.67 | 65.2 | 0.118 | O-VIII |
| 6.67 | 67.3 | 0.122 | O-VIII |
| 6.75 | 76.6 | 0.134 | O-III |
| 6.79 | 82.0 | 0.139 | O-III |
| 6.87 | 92.3 | 0.154 | O-I |

**Table S1: Oxygen content, $T_c$ and doping of samples used in this study.**

**MEASUREMENT OF THE SOUND VELOCITY IN HIGH MAGNETIC FIELD**

Samples were measured in static fields at the LNCMI-Grenoble or at the HFML-Nijmegen, and in pulsed fields at the LNCMI-Toulouse. The magnetic field $H$ was applied along the $c$-axis of the orthorhombic structure, perpendicular to the $CuO_2$ planes.

Sound velocity variation $\Delta v_s/v_s$ was measured using a standard pulse-echo technique[6]. $LiNbO_3$ transducers were glued on the polished surfaces of the sample using epoxy. Measurements were performed between 100 MHz and 400 MHz. In this study we measured the sound velocity of the longitudinal mode propagating along the $b$-axis, $c_{22}$, and the transverse mode propagating along the $a$-axis with polarization along the $b$-axis, $c_{66}$. We use the Voigt notation: $\varepsilon_2=\varepsilon_{yy}$ and $\varepsilon_6=\varepsilon_{xy}$ where $\varepsilon$ is the strain.

## 2. Superconductivity and elastic constant in zero magnetic field

It has been shown in cuprates that the impact of superconductivity on the lattice compressibility is twofold[7]. The first effect is a negative jump at the superconducting transition temperature $T_c$. This jump is linked to the jump in the isobaric heat capacity $\Delta C_p(T_c)$ through the Ehrenfest relation:

$$\Delta v_{ii}/v_{ii} = \frac{\Delta C_p(T_c)T_c}{2c_{ii}}\left(\frac{d\ln T_c}{d\varepsilon_{ii}}\right)^2 \qquad (1)$$

where the first factor is a ratio of the condensation energy and the elastic energy, and the second factor involves the strain dependence of $T_c$ (ref. 8). In Fig. S1a we show this negative jump in the sound velocity variation $\Delta v_s/v_s(T)$ of the longitudinal mode $c_{22}$ measured in a sample with $p = 0.11$. To extract the jump from the raw data, a background is subtracted, using a fit that models the lattice hardening due to anharmonic effect[9]. The background-subtracted data are shown in Fig. S1b. We extract the amplitude of the jump at $T_c$: $\Delta c_{22}/c_{22} = 6.2 \ 10^{-5} \pm 0.2 \ 10^{-5}$, which turns into $|dT_c/d\varepsilon_2|=348$ K using $\Delta C_p(T_c)/T_c=14$ mJ / mol / $K^2$ for heat capacity measurements[10] and $c_{22} = 268$ GPa for the elastic constant[11]. This value is in good agreement with values found in YBCO with thermal expansion[12].

The second effect is a progressive hardening below $T_c$. The lattice inside the superconducting state is harder than it would be in the absence of superconductivity. This was noticed earlier in $La_{2-x}Sr_xCuO_4$ (LSCO) and attributed to the positive curvature of strain dependence of the condensation energy[7].

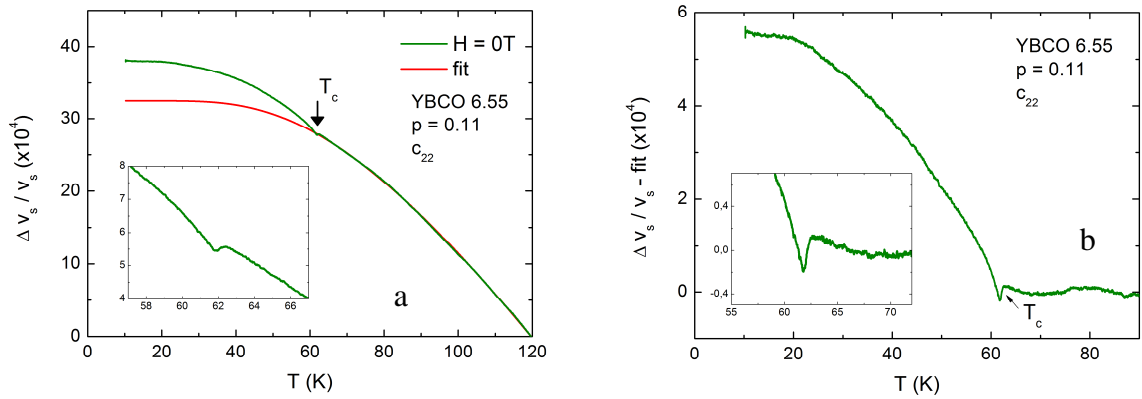

**Fig. S1: Impact of superconductivity on the lattice in cuprates.**

(a) The green curve is the relative change in the sound velocity associated with the $c_{22}$ mode as a function of temperature in zero magnetic field. The inset is a zoom of the region close to the superconducting transition temperature $T_c$ = 62.0 K. The red line is a fit to the data of the form $c - \frac{s}{\exp\left(\frac{t}{T}\right) - 1}$, which phenomelogically describes the hardening of the lattice due to anharmonic contribution to the lattice potential[9]. (b) Background-subtracted data showing the superconducting contribution to the elastic constant.

### 3. Sound velocity and charge order transition in high magnetic fields

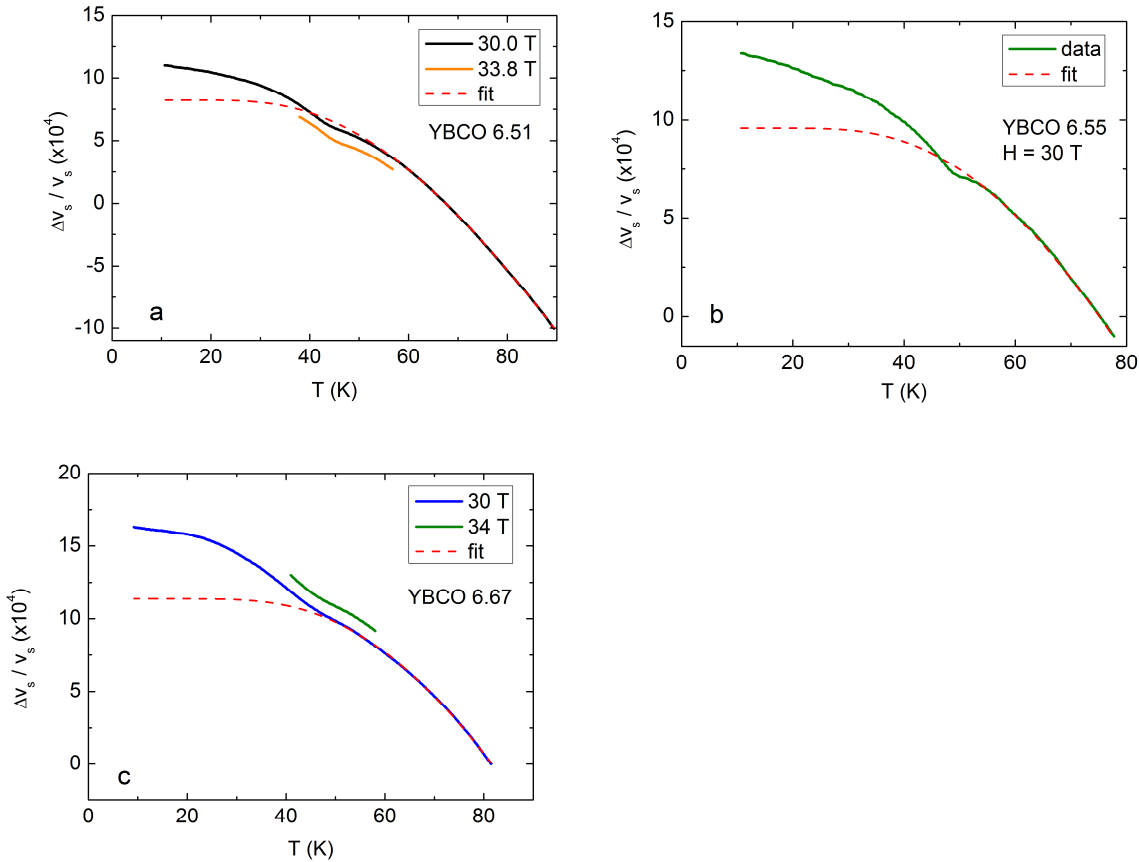

**Fig. S2: Sound velocity as a function of temperature in high magnetic fields.**

Sound velocity of the $c_{22}$ mode as a function of temperature measured in YBCO at (a) $p = 0.106$, (b) $p = 0.110$, and (c) $p = 0.122$ in high magnetic fields. The red dashed line is a fit to the data of the form $c - \frac{s}{\exp\left(\frac{t}{T}\right) - 1}$ performed above $T_{CO}$, that models the anharmonic lattice contribution to the elastic constant[9]. The fit shown here are subtracted to the data, and the resulting curves are shown in Fig. 1.

Let's evaluate the strain dependence of the charge order critical temperature $T_{CO}$ using the Ehrenfest formula in YBCO 6.55 ($p = 0.11$). While there is no specific heat data at the charge order transition temperature, we can estimate the specific heat jump at the charge order transition from the field dependence of the specific heat from Marcenat *et al.* which is at most 1 mJ/mol/K$^2$ (Ref. 10). From this value we get $|dT_{CO}/d\varepsilon_2| = 1460$ K, significantly higher than the strain dependence of the superconducting $T_c$.

In order to evaluate the hydrostatic pressure dependence of $T_{CO}$ we make the following approximation:

$$\frac{dT_{CO}}{dP} = \sum_i \frac{dT_{CO}}{dP_i} = \sum_i \sum_j s_{ij} \frac{dT_{CO}}{d\varepsilon_i} \sim \sum_j s_{2j} \frac{dT_{CO}}{d\varepsilon_2}$$

with $s_{ij}$ the compliance tensor, determined using measurements from Ref. 11. This leads to $|dT_{CO}/dP|$ = 1.7 K/GPa. Given the quasi-two dimensional nature of the cuprates and the fact that high field charge order develops along the $b$-axis, this approximation appears reasonable.

## 4. Sound velocity and the vortex lattice

In presence of a magnetic field lower than the upper critical field $H_{c2}$, one must take into account the elastic properties of the flux line lattice. The vortex lattice has 3 characteristic moduli: a tilt $c^v{}_{44}$, shear $c^v{}_{66}$ and compression moduli $c^v{}_{11}$. For $H /\!/ c$ and a sound wave propagation in the plane, the sound velocity probes $c^v{}_{11}$. For $H >> H_{c1}$, $\kappa >> 1$ and assuming negligible dispersion effect at low enough frequency[13] we have $c^v{}_{11} \sim H^2/4\pi$.

The vortex lattice when pinned to the crystal lattice has a contribution $\Delta c^v$ to the sound velocity $v_s$: $\rho v_s^2 = c_{ij} + \Delta c^v$, with $\rho$ the density of the system, and $c_{ij}$ the crystal elastic constant. This contribution is strongly influenced by the pining energy of the vortex lattice. In the early 90's Pankert *et al.*, developed and tested a phenomenological model based on thermally assisted flux-flow (TAFF) to describe the influence of the vortex lattice on the sound velocity in cuprates[14]. Within this model, the vortex lattice contribution to the sound velocity is:

$$\Delta c^v = c_{ii}^v \frac{\omega^2}{\omega^2 + (c_{ii}^v \Gamma k^2)^2}$$

With $\omega$ the sound wave angular frequency, $k$ the sound wave vector, and $\Gamma(T, H)$ a phenomenological relaxation rate, related to the resistive behavior caused by vortex motion in the TAFF regime. This model has been shown to successfully describe sound velocity measurements in $Bi_2Sr_2Ca_{n-1}Cu_nO_{2n+4+x}$ (Ref. 15) and in LSCO (Ref. 16). In Fig. S3 we use this model to illustrate the behavior of $\Delta c^v$ as a function of magnetic field, for a set of parameters determined in LSCO (Ref. 16). As temperature is reduced (red to blue), the depinning of the vortex lattice occurs at higher magnetic fields. When pining is lost, the vortex contribution to the sound velocity is also lost, and a step-like decrease is observed in the measured sound velocity. Such decrease is observed experimentally (see Fig. S3 right and Fig. S4), and we take the midpoint of this step-like feature to determine the vortex transition field $H_v$.

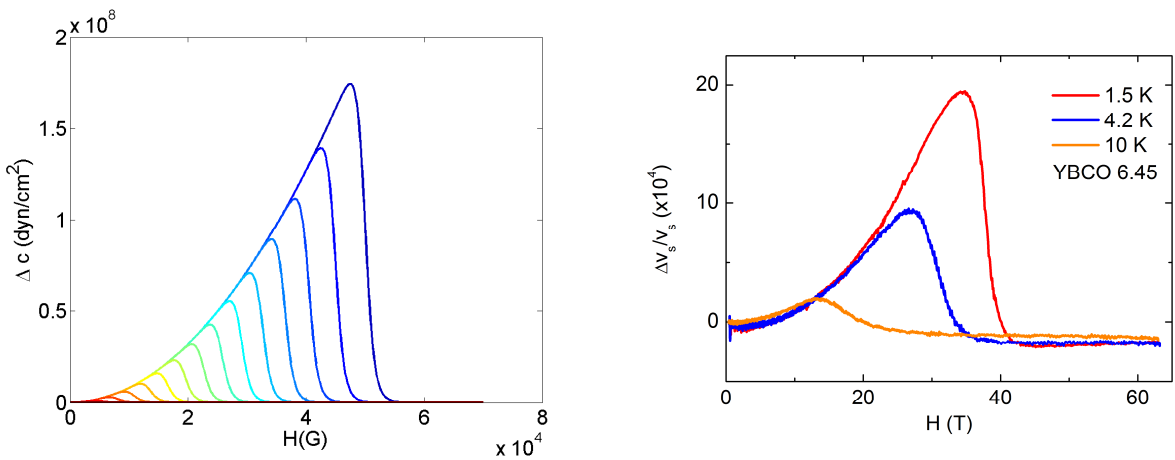

**Supplementary Figure S3: Impact of vortex lattice on the sound velocity.**

**Left panel**: $\Delta c^v$ computed as a function of magnetic field at different temperatures (5 K in blue to 20 K in red with 1K steps). The computation was performed using parameters found by Hanaguri *et al.,* in LSCO, for *H // c* and *k // a*, shown in Table 4 of ref. 16. As long as the vortex lattice is pined, it gives a $H^2$ contribution to the sound velocity. When depinning occurs, this contribution is lost, and the sound velocity decreases. **Right panel**: Sound velocity change as a function of magnetic field in YBCO 6.45 (*p* = 0.071) at different temperatures. At this particular doping no charge order is observed, and contribution of the superconducting order parameter to the elastic constant is small. Those conditions allow to highlight the vortex lattice contribution. A $H^2$ dependence is observed at low fields until a large drop occurs due to the loss of the vortex lattice elastic moduli contribution to the sound velocity.

# 5. Field dependence of the sound velocity ($c_{22}$ mode) and $H$-$T$ phase diagrams in YBCO at various doping

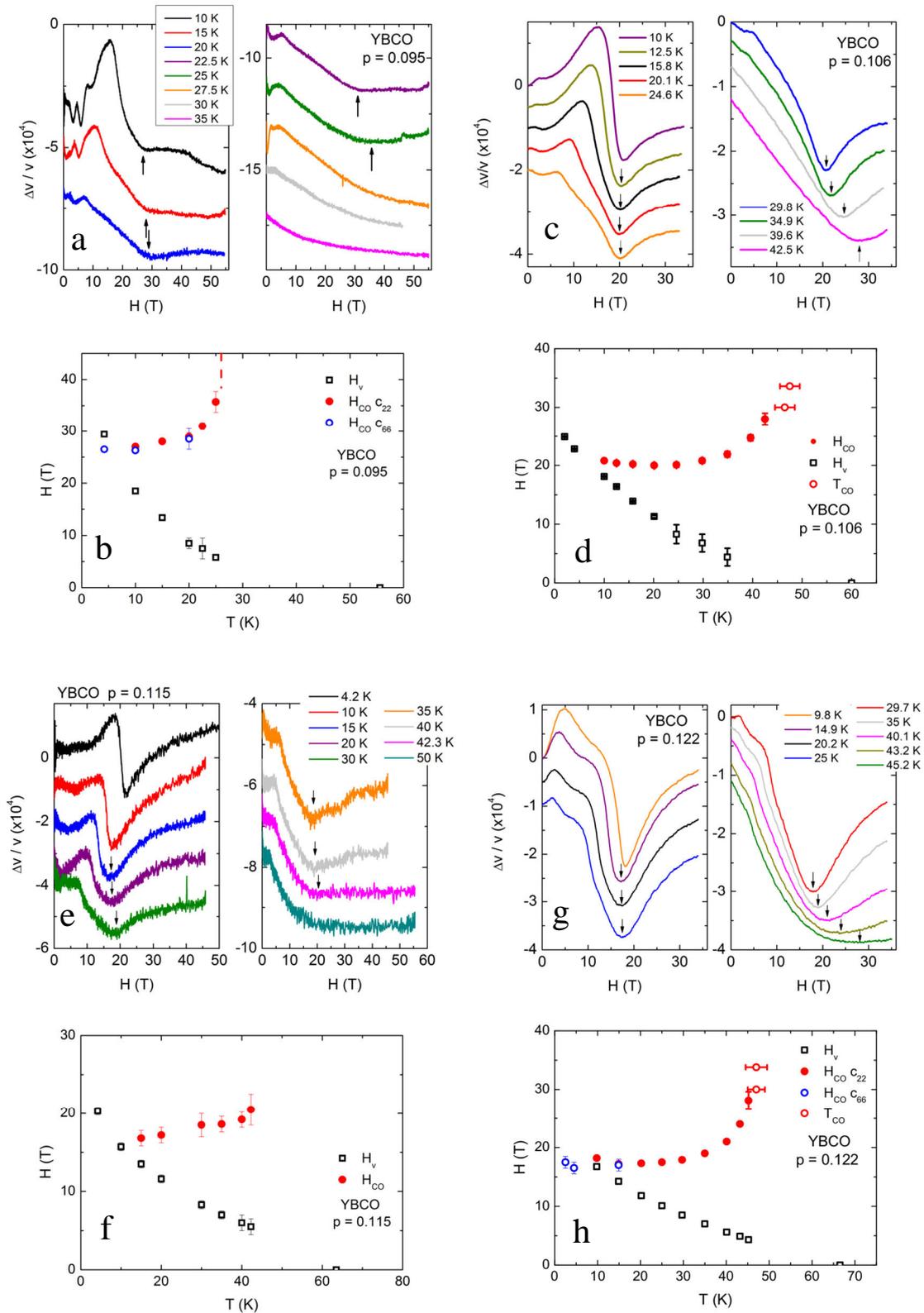

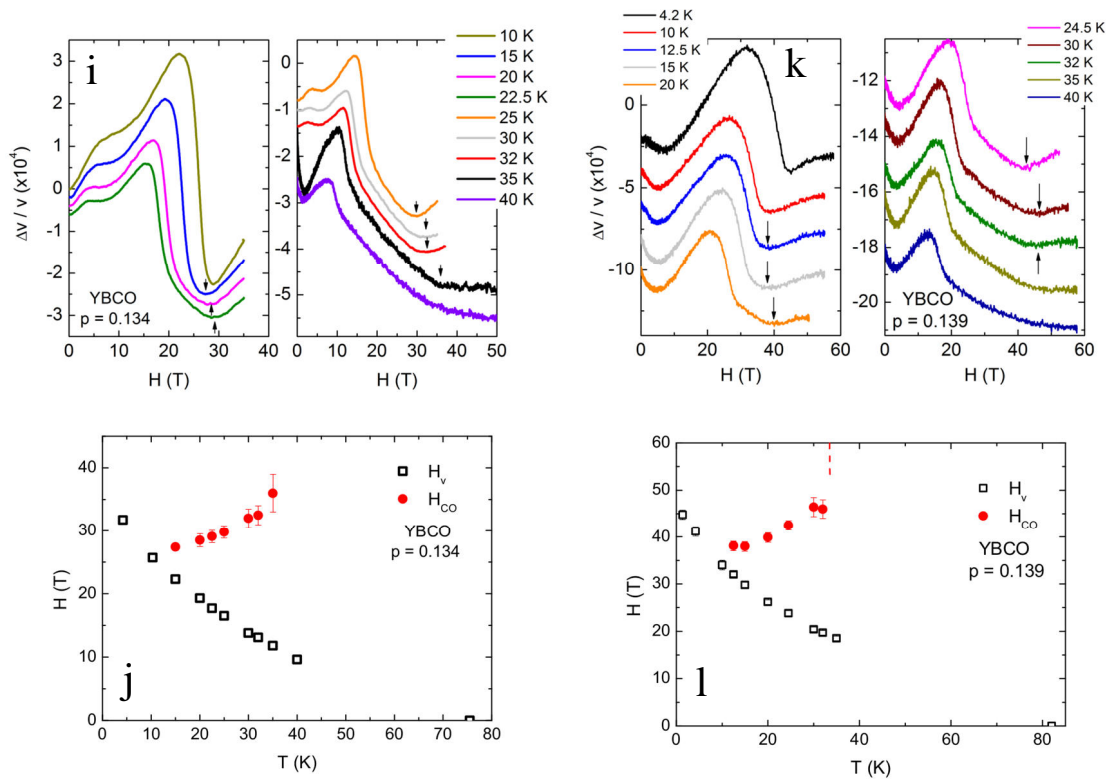

**Fig. S4: Sound velocity as a function of magnetic field and H-T phase diagrams for all doping.**

This figure is organized as follows: for each doping, sound velocity of the $c_{22}$ mode measured as a function of magnetic field for different temperatures is shown in the split upper panel. Arrows indicate $H_{CO}$, which is then reported as full red circles in a *H-T* diagram in the lower panel. $H_V$ is the vortex transition field as detected by ultrasound. (a) and (b): data for $p = 0.095$. Blue open circles show $H_{CO}$ as determined with measurements of $c_{66}(H)$, shown in Fig S5. (c) and (d) data for $p = 0.106$. In addition to $H_{CO}$, the *H-T* diagram features $T_{CO}$ in open red circles, as determined with temperature sweeps at constant magnetic field, shown in Fig. 1. (e) and (f) data for $p = 0.115$. (g) and (h) data for $p = 0.12$, also shown in Fig. 2. Additional points obtained with $c_{66}$ are included in the *H-T* diagram. (i) and (j) data for $p = 0.134$. (k) and (l): data for $p = 0.139$. The dashed lines in panel b and panel l indicate the temperature $T_{CO}$ above which no anomaly is detected in the field dependence of $c_{22}$.

## 6. Transverse mode $c_{66}(H)$

In Fig.S5, we show the field dependence of $c_{66}$ measured at $p = 0.095$ (Fig. S5a) and $p = 0.122$ (fig. S5b). The behaviour is very similar to that reported in ref. 17. At low field, a minimum appears in $c_{66}(H)$, which most likely finds its origin in the vortex physics. At higher fields, a kink followed by a hardening appears at the charge order transition (indicated by an arrow) for $p = 0.122$ and $p = 0.095$. This behavior persists even for temperatures where $H_{CO} < H_v$, allowing to determine $H_{CO}$ down to the lowest temperatures, and to complete the phase diagrams of Fig. S4b. Above 15 – 20 K or so the transition is smeared out in $c_{66}$ and it is more accurately determined in $c_{22}$, where the signal is 10 times larger. In the temperature range where the transition can be observed in both $c_{22}$ and $c_{66}$, we find good agreement between the two measurements within error bars, as shown in Fig S4b and S4h. The $c_{66}$ anomaly associated with the charge order transition observed in YBCO $p = 0.095$ and $p = 0.122$, is absent at $p = 0.154$ as shown in Fig. S5c. This demonstrates that charge order is not present at $p = 0.154$ down to 4.2 K and up to 66 T.

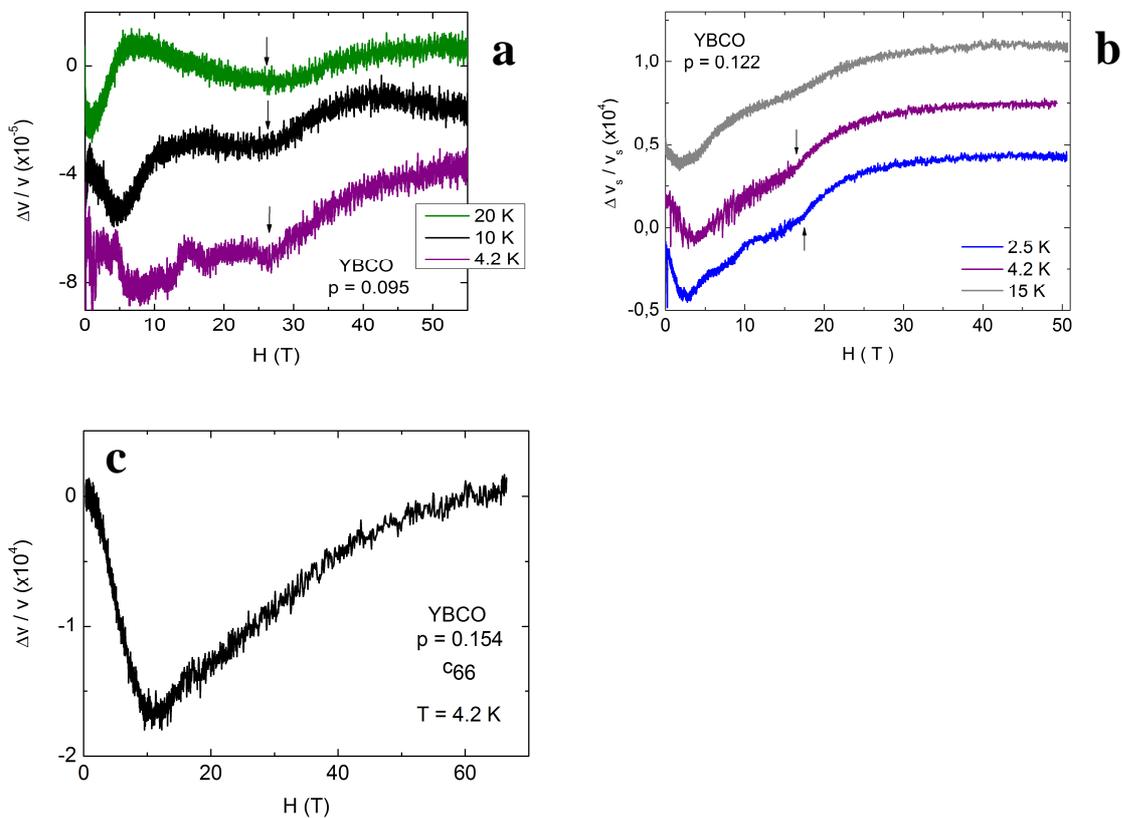

**Fig. S5: Transverse mode $c_{66}(H)$ at different dopings.**

Sound velocity of the transverse mode $c_{66}$ as function of magnetic field, at different temperatures in YBCO at (a) $p = 0.095$, (b) $p = 0.122$, and (c) $p = 0.154$.

## 7. Comparison of different length scales

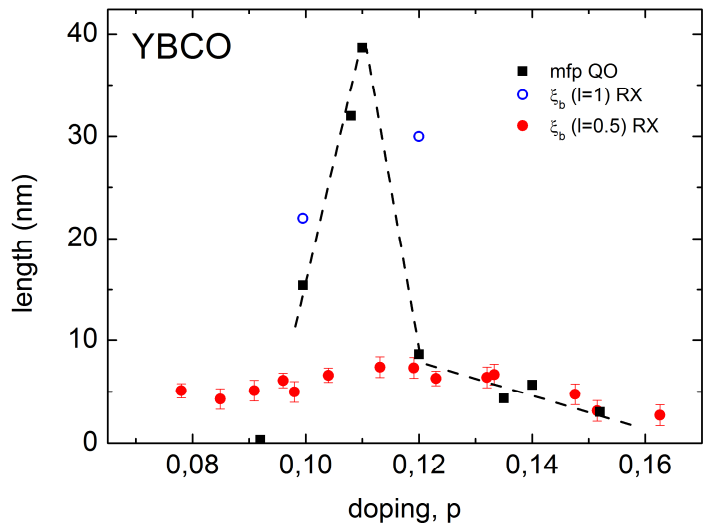

**Fig. S6: Comparison of charge orders correlation lengths and quasiparticle mean free path.**

Doping dependence of the mean free path deduced from quantum oscillation measurements (black squares), of the correlation length along the *b*-axis for the 2D charge order at *H* = 0 (red circles) and of the correlation length along the *b*-axis for the 3D charge order in high fields (blue open circles). Quasiparticle mean free path deduced from quantum oscillation experiments reported in Ref. 18,19 and 20. 2D charge order correlation length measured at $T_c$ as reported in Ref. 2 and 21. 3D charge order correlation length in finite magnetic field as reported in Ref. 22, 23.

## 8. Comments on the de Broglie wavelength – charge order correlation length comparison

In the main text the de Broglie wave length $\xi_{th}$ is evaluated using the Fermi velocity $v_F$ deduced from quantum oscillation measurements of the reconstructed Fermi surface. Ideally our analysis would require the use of the Fermi velocity of the Fermi surface that exists before the onset of Fermi surface reconstruction. However, in YBCO the reconstruction involves a folding process, as suggested in most scenarios[24]. Consequently, the Fermi velocity should be similar before and after reconstruction. Moreover, the Fermi velocity deduced from quantum oscillation measurements is an averaged velocity over the entire cyclotron orbit. Yet, only $v_F$ measured near those regions of reciprocal space where Fermi surface segments are connected through the charge order wave vector (namely the anti-nodal regions) should be considered for the comparison of de Broglie wave length and charge order correlation length. Nonetheless, we expect $v_F$ near the antinodes to be smaller than the averaged $v_F$ deduced from quantum oscillations such that the condition $\xi_{CO} \geq \xi_{th}$ (see Fig. 6b) is even more fulfilled.